\documentclass[aps,prb,floats,preprintnumbers,superscriptaddress]{revtex4}
\usepackage{amsmath}
\usepackage{amsfonts}
\usepackage{amssymb}
\usepackage{color}
\usepackage{bm}
\usepackage{cleveref}
\usepackage{bm}
\newcommand{\beq}{\begin{equation}}
\newcommand{\eeq}{\end{equation}}
\newcommand{\beqa}{\begin{eqnarray}}
\newcommand{\eeqa}{\end{eqnarray}}
\newcommand{\bs}{\boldsymbol}

\usepackage{graphicx,amssymb,amsmath
}
\newcommand{\bb}[1]{\mathbf{#1}}

\newcommand{\mth}[1]{\mathcal{#1}}

\newcommand{\vf}{v_\text{F} }
\newcommand{\intt}{\int\displaylimits}

\begin{document}
\bibliographystyle{naturemag}

\title{Fingerprints of the conformal anomaly on the thermoelectric transport in Dirac and Weyl semimetals}
\author{Vicente Arjona$^{1}$, Maxim N. Chernodub$^{2,3}$,  and Mar\'ia A.H. Vozmediano$^{}$}
\email{${1}$vicente.arjona@hotmail.com, ${2}$maxim.chernodub@gmail.com, ${3}$vozmediano@icmm.csic.es}
\affiliation{
${1}$ Instituto de Ciencia de Materiales de Madrid, C/ Sor Juana In\'es de la Cruz 3, 
Cantoblanco, 28049 Madrid, Spain\\
${^2}$Institut Denis Poisson UMR 7013, Universit\'e de Tours, 37200 France\\
$^{3}$Laboratory of Physics of Living Matter, Far Eastern Federal University, Sukhanova 8, Vladivostok, 690950, Russia}

\begin{abstract}

A quantum anomaly arises when a symmetry of the classical action can not survive quantization. The physical consequences of having quantum anomalies were first explored in the construction of quantum field theory to describe elementary particles and played an important role in grand unification and string theory. Nowadays, the interest on anomalies and anomaly related transport has shifted to emergent condensed matter systems which support low energy descriptions akin to their QFT partners. Dirac and Weyl semimetals are 3D crystals having band crossings near the Fermi surface. Their low energy quasiparticles are described by a massless Dirac Hamiltonian sharing all the properties of their high energy counterparts. After an intense and successful analysis of the consequences of the chiral anomaly on magneto-electric transport, the interest has shifted to gravitational effects, in particular these of the mixed axial-gravitational anomaly. These phenomena involve thermo-electric measurements in magnetic field. A less known quantum anomaly, the conformal anomaly, also related to metric deformations, has been recently shown to give rise to a special contribution to the Nernst signal which remains finite at zero temperature and chemical potential. In this work we provide distinctive signatures for the experimental confirmation of this unexpected signal.

\end{abstract}

\maketitle

\section{Introduction}

According to the Noether theorem, a continuum symmetry of a classical action gives rise to conserved currents and charges. Energy-momentum or angular momentum are associated to space-time translation and rotations, while internal phase rotations of complex fields give rise to electric, color, or other conserved currents and charges. A quantum anomaly arises when a symmetry of the classical action can not survive quantization. Normally the origin of these anomalies occurs in the presence of interactions and originates in the difficulties encountered when  the classical currents are substituted by local operators \cite{Bert96,Hol93}. The physical consequences of having quantum anomalies were first explored in  the construction of quantum field theory (QFT) to describe elementary particles \cite{Bert96}   and played an important role in grand unification and string theory. Nowadays, the interest on anomalies and anomaly--related transport  has shifted to emergent condensed matter systems which support low energy descriptions akin to their QFT partners \cite{Dima14,Karl14}. Dirac and Weyl semimetals are 3D crystals having band crossings near the Fermi surface. Their low energy quasiparticles are described by a massless Dirac Hamiltonian sharing all the properties of their high energy partners. After an intense and successful analysis of the consequences of the chiral anomaly on magneto-electric transport \cite{KK13,XKetal15,Lietal15,ZXetal16}, the interest has shifted to gravitational effects, in particular these of the mixed axial-gravitational anomaly \cite{Getal17,Setal18}. These phenomena involve thermo-electric measurements in magnetic field.

The way gravity appears in material physics can be traced back to the problem of defining thermodynamic equilibrium in curved backgrounds \cite{TE30} which culminated with the Luttinger theory of thermal transport \cite{Lut64}. The difficulty to find a local source for thermal (energy) currents was solved by introducing a (may be fictitious) gravitational field whose gradient plays the role of the electric field in the electromagnetic transport. This is a very natural choice in QFT where the stress tensor is the response to variations of the metric: $T^{\mu\nu}\sim\delta S/\delta g_{\mu\nu}$.  

Anomalies are generally due to vacuum fluctuations and often the induced transport responses persist at zero temperature and chemical potential. In a recent publication \cite{LLetal17,WMetal18,CCV18} a less known quantum anomaly, the conformal anomaly, also related to metric deformations, has been shown to give rise to a special contribution to the Nernst signal which remains  finite at zero temperature and chemical potential.  

Thermoelectric transport is a topic of major interest in technology and a very important tool to analyze the electronic properties of materials. From the early research it was known that semiconductors and semimetals are the best candidates to generate large figures of merit in thermopower, with bismuth, an almost compensated semimetal, holding the record for metallic compounds \cite{BA16}. Dirac and Weyl semimetals belong naturally to the family of good thermo-electric materials and their thermoelectric properties are now at the center of interest in experimental and theoretical research  \cite{LLF14,SGT16,FZB17,SMetal17,GMSS17,LLetal17,YYetal18,WMetal18,GSetal18,SK18,SMetal18}. Giant values of the anomalous Nernst effect are been systematically  reported in the newly-discovered magnetic Weyl semimetals. Most of the works deal with the anomalous Nernst and  Hall effects (transverse transport in the absence of external applied magnetic fields) associated to the non trivial Berry phase of the materials \cite{FZB17,LLetal17,MM18,SMetal18,YYetal18}. The corresponding anomalous coefficients are normally obtained using a semiclassical Boltzmann approach following the work in  \cite{XY06}. 

The regime of zero temperature and chemical potential where the unusual prediction in \cite{CCV18} of a non--zero transport coefficient lies, prevents the use of Boltzmann approach and the comparison with existing results.  We present a Kubo calculation of the thermoelectric coefficient of the massless Dirac system in a magnetic field  at zero chemical potential and zero temperature.  The result coincides with the one obtained in \cite{CCV18} putting on firmer grounds the anomaly--related transport phenomena \cite{Karl14,Dima14} in the Dirac matter. An extension of the analysis to finite temperature and chemical potential confirms the robustness of the anomaly induced contribution and provides distinctive signatures for the experimental confirmation of this unexpected signal. The result is independent of the Berry curvature and hence it is common to Dirac and Weyl semimetals in the low $T$ and low $\mu$ limit.

\section{Dirac and Weyl semimetals}
\label{sec_dirac}
 \begin{figure}
\centering
\includegraphics[width=0.5\columnwidth]{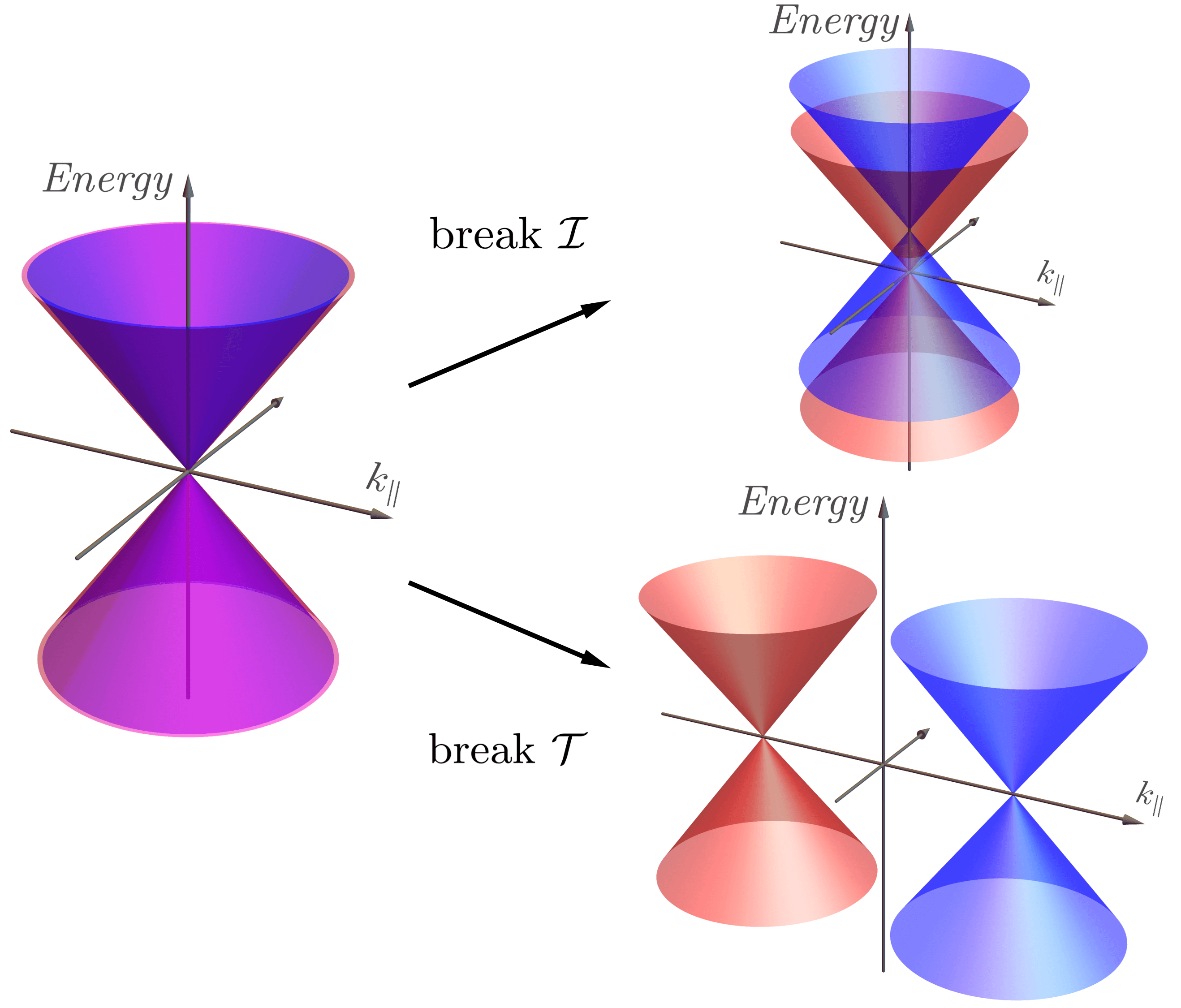}
\caption{ Schematic dispersion relation of Dirac (left) and Weyl semimetals (right).}
\label{fig_dirac}
\end{figure}
The low energy excitations around a non-trivial band crossing of a Dirac semimetal are described by the massless Dirac equation in three space dimensions. In the Weyl basis for the Dirac matrices, the Hamiltonian splits into two Weyl nodes (two dimensional spinors) of definite chirality. 
The low energy Hamiltonian around the Dirac point can be written as $H(k)=s{\vec\sigma}\cdot{\vec k}$ where $s=\pm$ is the chirality. Each chirality acts as a monopole of  Berry curvature of charge $s$. In the material realizations, discrete symmetries associated to the crystal lattice plus standard inversion ${\cal I}$ and time reversal ${\cal T}$ play a crucial role. Depending of the symmetry of the crystal - and on the spin--orbit coupling -  we find Dirac or Weyl semimetals schematically shown in Fig.~\ref{fig_dirac}.  In the first class, the  two chiralities are superimposed in momentum space and a mass term can arise mixing the two chiralities, unless the band crossing is protected by crystal symmetries. Examples of symmetry protected Dirac semimetals are $Cd_3As$, $Na_3Bi$ \cite{AMV18}. In these materials time reversal symmetry is unbroken and the Hall conductivity is zero. In the Weyl semimetals to two chiralities are separated in momentum or energy (see Fig.~\ref{fig_dirac}). This separation necessarily breaks either ${\cal I}$ or ${\cal T}$ and the Berry monopole makes the Weyl points very robust against perturbations. A gap can open only by merging the two chiralities. 

Details of the material realizations can be found in many good reviews \cite{HPV12,JXH16,AMV18}. What is important for our work is to realize that the conformal invariance of the classical system implies that no dimension-full parameter enters in the description. Our model will then generically be that of a massless Dirac semimetal as used in ref. \cite{CCV18}. We will perform the calculation for each Weyl fermion and make sure that no cancellation occurs due to the opposite chiralities. Once this is confirmed, the result will equally apply to Weyl semimetals which, eventually, can receive additional contributions from the separation of the Weyl points (a dimension-full parameter).

\section{Thermoelectric transport: Kubo calculation}
\label{sec_calculation}
The response of an electronic system  to a background electric field ${\bb E}$ and temperature gradient ${\bb \nabla}T$  is parametrized as,
\beqa
J^i&=e^2L_{11}^{ij}E_j+eL_{12}^{ij}\nabla_j T \\\nonumber
J_\epsilon^i&=eL_{21}^{ij}E_j+L_{22}^{ij}\nabla_j T.
\label{eq_main}
\eeqa
where the coefficients $L_{ab}^{ij}$, are related to the standard thermoelectric conductivities: 
\beq
{\bb J}=\hat\sigma\cdot {\bs \nabla V}+\hat\alpha ({-\bs \nabla T}),
\eeq
by $\sigma^{ij}=L_{11}^{ij}/e^2$  (electrical  conductivity), and  $\alpha^{ij}=eL_{12}^{ij}$ (thermopower).

The expression obtained in ref.~\cite{CCV18} for a single Dirac cone in three dimensions from the conformal anomaly (in the geometry $B_z$, $\nabla_yT$) was
\beq
J^x = \frac{e^2 v_F B}{18 \pi^2  \hbar}\left(\frac{ \nabla_y T}{T}\right) \, ,
\label{eq:us}
\eeq
from where we extract the thermoelectric coefficient:
\beq
\chi^{xy}=\frac{e^2 v_F B}{18 \pi^2  \hbar }  .
\label{eq_chi}
\eeq
This coefficient is related with the standard definition in Equation~\eqref{eq_main} by $\alpha^{ij}=\chi^{ij}/T$. 
In what follows we will present a standard Kubo formula calculation of this thermoelectric tensor. The calculation is straightforward but lengthy. We will sketch the main aspects and provide extensive information and details in the supplementary material.

\subsection{Kubo formula for the thermoelectric tensor}
\label{sec_Kubo}

In linear response theory \cite{GV05}, when the action of a system is perturbed by a local source $F(t)$ which couples to an observable $ B$ as $H_F={\bf F}\cdot{\bf B}$, the change in the expectation value of any operator $A$ is assumed to be linear in the perturbing source:
$\delta \langle  A^i(t) \rangle=\int dt' \chi^{ij}(t,t')F_j(t')$,
and the response function $\chi^{ij}$ is given by the Kubo formula: 
\beq 
 \chi^{ij}(t, t')=  -\frac{i}{\hbar} \int\displaylimits_{-\infty}^\infty \! \text{d}t' 	\, \Theta(t-t') \langle \left[ \hat{A^i} (t) , \hat{B^j} (t') \right] \rangle_0 F(t'),   
\eeq 
where $\hat{A}, \hat{B}$ are operators in the interaction picture representation and $\Theta(x)$ the Heaviside function.

The problem of using a statistical variable (such as the temperature) as a (local) source coupling to an energy current was  solved by Luttinger in \cite{Lut64}. Based on previous analyses by Tolman and Ehrenfest trying to define thermal equilibrium in a curved space \cite{TE30}, he proposed the gravitational potential $\Phi$ as the local source of thermal (energy) current $J_\epsilon^i=T^{0i}$.  Physically, the observation in \cite{TE30} was that a temperature perturbation which moves a system out of equilibrium can be compensated by a variation in the gravitational potential such that, in equilibrium, (we take the speed of light c = 1),
\beq
\nabla\Phi+\frac{\nabla T}{T}=0.
\label{eq_Lutt}
\eeq
For small deviations from flat space the gravitational potential is proportional to the zero-zero component of the metric $\Phi\sim g^{00}$ which couples to the energy density $T^{00}$. The perturbed Hamiltonian to be used in the linear response formalism is:
\beq 
H_{pert} (t) = \Theta(t-t_0)  T^{00} (t) g_{00} (t).
\label{eqn:per}
\eeq 
The electric current generated by this perturbation in linear response is: 
\beq 
\langle J^{i} \rangle (t,\bb r) = \int\displaylimits_{-\infty}^\infty \! \text{d}t' \text{d}\bb r ' \,  \underbrace{\left\{\frac{-i}{\hbar } \Theta(t-t') \langle \left[ J^{i} (t, \bb r) , T^{00} (t', \bb r ') \right]\rangle   \right\}}_{\chi^i (t-t', \bb r - \bb r ')} g_{00} (t', \bb r' ) ,
\label{eq:J}
\eeq 
where the system is assumed to be invariant under time and spatial translations.

Next we use the conservation law of the energy-momentum tensor $T^{\mu \nu} $: 
$\partial_0 T^{00} (t,\bb r)  + v_\text{F} \partial_i T^{0i}(t,\bb r) = 0 $,
in order to get the gradient of the gravitational potential which represents the thermal perturbation from \Cref{eq_Lutt} (notice that we have introduced the Fermi velocity in the spatial part of the metric to adapt the calculation to the case of Dirac semimetals). We then have
\beq
T^{00} (t,\bb r) = \int\displaylimits_{-\infty}^{t} \! \text{d}t' \left( -\vf \partial_i T^{0i} (t', \bb r)  \right) , 
\label{eqn:con}
\eeq 
where we have used that the system is unperturbed at $t=-\infty$. 
Introducing \Cref{eqn:con} in \Cref{eq:J} and integrating by parts, we get: 
\beq 
\langle J^{i} \rangle (t,\bb r) = \int\displaylimits_{-\infty}^\infty \! \text{d}t' \text{d}\bb r ' \int\displaylimits_{-\infty}^{t'} \text{d}t'' \,  \left\{\frac{-i \vf }{\hbar } \Theta(t-t') \langle \left[ J^{i} (t, \bb r) , T^{0j} (t'', \bb r ') \right]\rangle   \right\} \partial_{j'} g_{00} (t', \bb r' ) .
\label{eqn:ahh}
\eeq 
\Cref{eqn:ahh} represents the electric current generated by a thermal gradient computed via Kubo formula. The   Fourier transform 
\beq 
\langle J^{i} \rangle (\omega, \bb q) =   \chi^{ij} (\omega, \bb q) ( i q_j ) g_{00} (\omega, \bb q) ,
\label{eqn:ec}
\eeq 
leads to the standard form of the response function: 
\beq
\chi^{ij} (\omega, \bb q) =  (2\pi)^3 \int\!  \text{d}t \,  e^{i \omega (t-t') } \! \intt_{-\infty}^{t'} \!\text{d}t'' \, \left \{ \frac{-i \vf }{\mathcal{V}\hbar} \Theta(t-t') \langle \left[ J^i (t, \bb q) , T^{0j} (t'', -\bb q) \right] \rangle \right\} ,
\label{eq:rf}
\eeq 
where $\mathcal{V}$ is the volume of the system.
This is the expression that we will compute for a Dirac semimetal in the presence of an external magnetic field.

\subsection{Highlights of the calculation}
\label{sec_calculo}
The Hamiltonian of a Dirac semimetal in an external magnetic field can be decomposed into
\beq 
H_s = s \vf  \sigma^i ( p_i + e A_i) , 
\eeq 
where $s$ describes the chirality ($s=\pm$) and $e$ is the charge of the electron ($e=|e|$). Choosing the Landau gauge $A_x= -B y$ to characterise a magnetic field in the $z$-direction, the spectrum of the Hamiltonian is: 
\beqa 
E_{k_z m s} & =& \text{sign}(m) \vf \left[ 2 e \hbar B \vert m \vert  + \hbar^2 k_z^2 \right]^{1/2}      \hspace{1cm}  m\in \mathbb{Z}, \; m\neq 0\;,
\\ 
E_{k_z 0 s} &=& - s \vf \hbar k_z .
\label{eq_LL}
\eeqa 
The two zeroth Landau levels have opposite chiralities and the rest are doubly degenerated. 
\begin{figure}
\centering
\includegraphics[width=0.5\columnwidth]{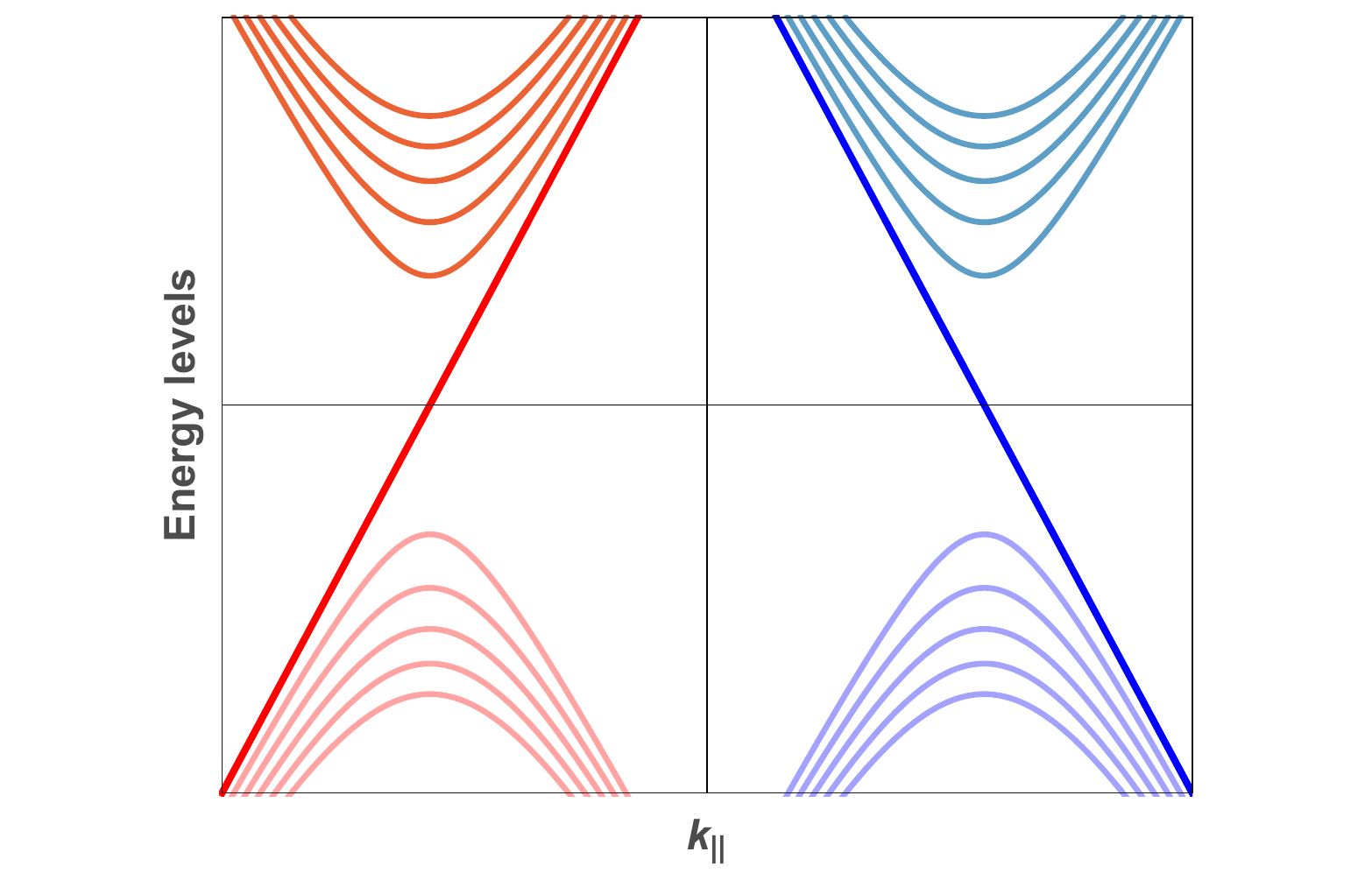} 
\caption{ Landau level structure of the two chiralities in a Weyl semimetal. In the Dirac semimetals the two images are superimposed.}
\label{fig_LL}
\end{figure} 
The energy levels  are sketched in Fig.~\ref{fig_LL}. The eigenvectors are: 
\beq 
\varphi_{\bb k m s } (\bb r) = \frac{1}{\sqrt{L_x L_z }} \frac{e^{i k_x x} e^{i k_z z}}{\sqrt{\alpha_{k_Z m s }^2+1}}e^{-(y-k_x l_B^2)^2/2l_B^2} \begin{pmatrix} 
 \frac{\alpha_{k_z m s}}{\sqrt{2^{M-1} (M -1)! \pi^{1/2} l_B } } H_{M - 1} \left[\frac{y-k_x l_B^2 }{l_B}\right] \\ \frac{1}{\sqrt{2^{M } M ! \pi^{1/2} l_B } } H_{M } \left[\frac{y-k_x l_B^2 }{l_B}\right]
\end{pmatrix} ,
\label{eqn:lw}
\eeq
with
\beq
\alpha_{k_z m s} =\frac{-\sqrt{2 e B \hbar \vert m \vert }}{E_{k_z m s}/s \vf - \hbar k_z } .
\eeq 
Capital letters refer to the absolute value of Landau levels, $H_m(x)$ are the Hermite polynomials, and the factor $\sqrt{\alpha_{k_z m s}^2 +1 } $ comes from the wave-function normalization. 
In the basis of the Landau levels, the current operators in \Cref{eq:rf} read:
\beq 
\hat{J}^x (t, \bb q) = \sum_{\bb k , m n} J^x_{\bb k m s, \bb k + \bb q n s } (+\bb q ) \hat a^\dagger_{\bb k m s} (t) \hat a_{\bb k + \bb q n s} (t)
\label{eqn:qq0}
\eeq   
\beq 
\hat{T}^{0y} (t'', -\bb q) =  \sum_{\bold{\boldsymbol{\kappa}} , \mu \nu}  T^{0y}_{\bold{\boldsymbol{\kappa}} \mu s,\bold{\boldsymbol{\kappa}}-\bb q \nu s} (-\bb q) \hat a^\dagger_{\bold{\boldsymbol{\kappa}} \mu s} (t'') \hat a_{\bold{\boldsymbol{\kappa}} - \bb q \nu s} (t''),
\label{eqn:qq1}
\eeq   
where the matrix elements are 
\beq 
J^x_{\bb k m s ,\bb k + \bb q n s } (\bb q) =\frac{1}{(2\pi)^{3/2}} \intt\! \text{d} y  \, e^{-i q_y y} \, s \vf e  \varphi^\ast_{\bb k m s} (y ) \sigma^x \varphi_{\bb k + \bb q n s} (y),
\label{eqn:qq2}
\eeq 
\begin{alignat}{2}
T^{0y}_{\bold{\boldsymbol{\kappa}} \mu s ,\bold{\boldsymbol{\kappa}} - \bb q  \nu s } (-\bb q ) &= \frac{1}{4} \frac{1}{(2\pi)^{3/2}} \intt \! \text{d} y' \, e^{+i q_y  y ' } \, \left[ \vf \varphi_{\bold{\boldsymbol{\kappa}} \mu s}^\ast (y ') \mathbb{I}  \{p_y \varphi_{\bold{\boldsymbol{\kappa}}-
\bb q  \nu s}  (y' )\}   -\vf \{p_y  \varphi_{\bold{\boldsymbol{\kappa}} \mu s}^\ast (y ') \}\mathbb{I}   \varphi_{\bold{\boldsymbol{\kappa}} -  \bb q  \nu s}  (y ' )    \right] \nonumber \\ 
                                          &+ \frac{1}{4} \frac{1}{(2\pi)^{3/2}} \intt \! \text{d}y' \, e^{+i q_y  y ' } \, \left[ \varphi^\ast_{\bold{\boldsymbol{\kappa}} \mu s}  (y' ) s \sigma^y (E_{\bold{\boldsymbol{\kappa}} \mu s}+E_{\bold{\boldsymbol{\kappa}}- \bb q \nu s}-2\mu)\varphi_{\bold{\boldsymbol{\kappa}}-\bb q  \nu s} (y' ) \right] .
\label{eqn:qq3}
\end{alignat}

Introducing  \cref{eqn:qq0,eqn:qq1,eqn:qq2,eqn:qq3} into the expression \Cref{eq:rf} the response function is given by: 
\beq 
\chi^{xy} (\omega, \bb q) = \lim_{\eta \to 0} \sum_{\bb k,m n} \frac{(2\pi)^3} { \mth V } \frac{ -i \vf \hbar \, J^x_{\bb k m s,\bb k+\bb q n s} (\bb q) T^{0y}_{\bb k+\bb q n s,\bb k m s} (-\bb q)}{ (E_{k_z m s} - E_{k_z + q_z  n s} +i \hbar \eta) (E_{k_z m s} - E_{k_z + q_z n s} +\hbar \omega +i \hbar \eta ) }  \left[  n_{\bb k m s} - n_{\bb k+\bb q n s} \right]  ,
\label{eqn:rf2}
\eeq 
where we have used the relation
\beq 
\langle [  \hat{a}_{\bb k m s}^\dagger  \hat{a}_{\bb k+\bb q n s} , \hat{a}_{\bold{\boldsymbol{\kappa}} \mu s}^\dagger  \hat{a}_{\bold{\boldsymbol{\kappa}}-\bb q \nu s}]\rangle = \delta_{\bb k,\bold{\boldsymbol{\kappa}}-\bb q} \delta_{m,\nu} \delta_{\bb k+\bb q,\bold{\boldsymbol{\kappa}} } \delta_{n,\mu} \left\{  n_{\bb k m s} - n_{\bb k+\bb q n s} ,\right\}.
\eeq 
and we have   added a factor $e^{-\eta t}$ in the time  integration  to guarantee the convergence of the time integrals.

As we see, there is a sum over energetically allowed transitions. 
In the limit $T\to 0$, the distribution function becomes a step function: $n_{\bb k m s } = \Theta(\mu - E_{k_z m s })$ and only transitions between positive and negative levels are allowed.   We have evaluated the  numerical value of the response function taking only into account the dominant contribution due to  transitions between  the lowest Landau levels that give: 
\beq 
\chi^{xy}\equiv\lim_{\bb q \to 0} \lim_{\omega \to 0} \chi^{xy} (\omega, \bb q ) = \frac{-1}{2(2\pi)^2}\frac{\vf e^2 B }{\hbar}.
\eeq  
The contribution of higher energy transitions change the numerical value by a factor of 2 approximately.   Finite temperature and chemical potential dependences will be discussed below.

\section{Experimental signatures. Prospects.}
\label{sec_res}
The calculation presented here is based on a three dimensional massless Dirac Hamiltonian that can describe Dirac or Weyl semimetals in the low energy regime. The result is valid irrespective of whether the opposite chiralities are superimposed (Dirac) or separated (Weyl) in momentum space. Additional contributions proportional to the separation of the Weyl nodes may arise in Weyl semimetals. We have been particularly careful to follow the chirality dependence of all the terms along the calculation to ensure that no cancellations occur. 
Placing the chemical potential at $\mu=0$, and adding the contributions of the two chiralities, the thermoelectric coefficient $\alpha^{xy}$ in \Cref{eq_main} is given by: 
\beq 
\alpha^{xy}=\frac{2}{T} \chi^{xy} = -\frac{e^2\vf  B}{4\pi^2 T\hbar } .
\eeq  
which coincides with the result in ref. \cite{CCV18} up to a numerical factor close to unity (the sign is a matter of convention). This is a remarkable result. First notice that, although we deal with thermally induced transport, the calculation has been done at zero temperature and the coefficient remains finite in the limit $T\to 0$ with $\nabla T/T $ being kept  finite. Second, this calculation is valid at the Dirac point at zero chemical potential where it captures the vacuum contribution from the quantum conformal anomaly. 
\begin{figure}
\centering
\includegraphics[width=0.6\columnwidth]{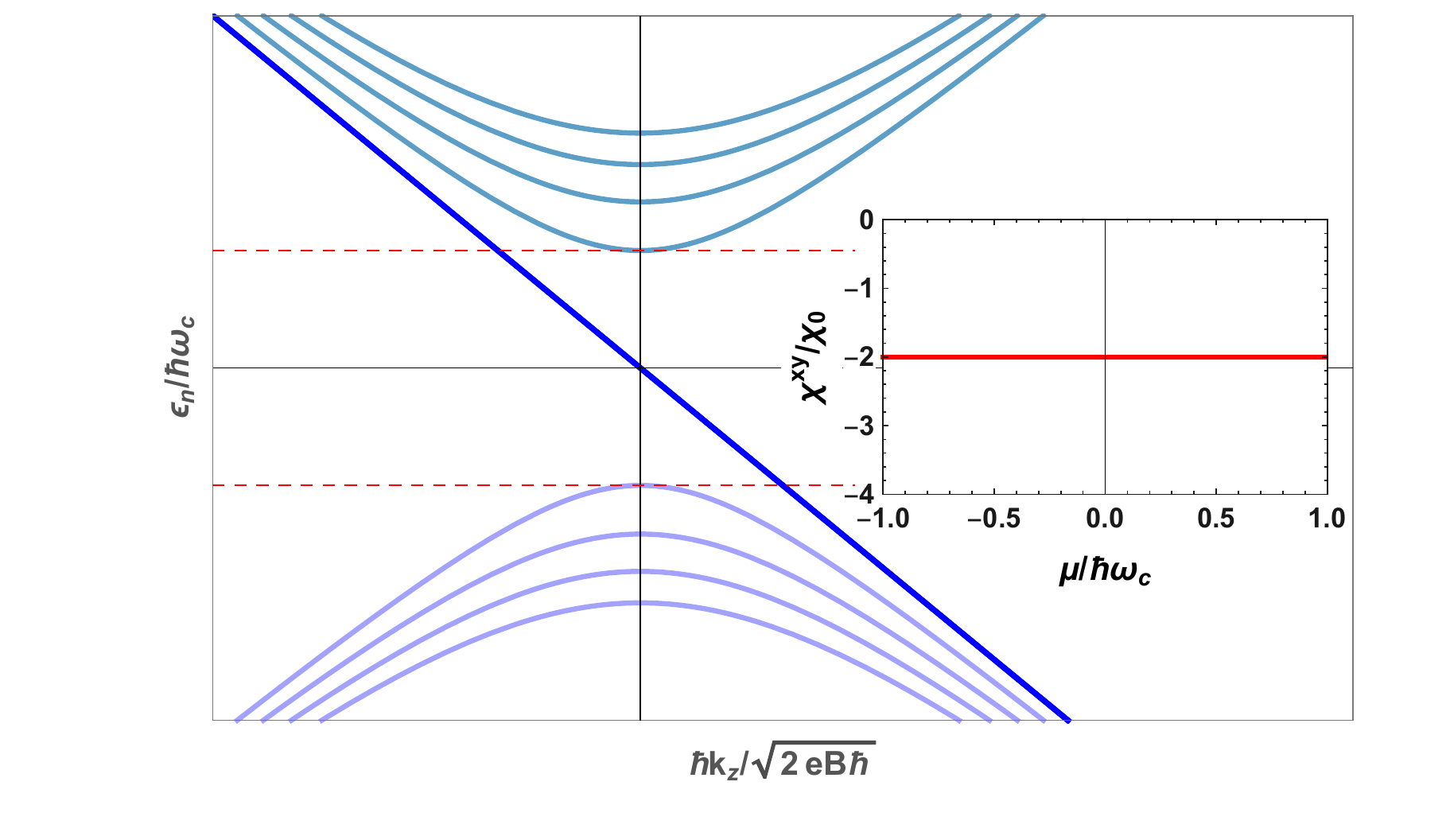} 
\caption{ Landau level structure of a single chirality in the Dirac semimetal. The inset shows the calculated thermoelectric coefficient as a function of the  the chemical potential in normalized units at $T=0$. The function has a constant value ($\chi^{xy}/\chi_0=-2$, where $\chi_0 = \vf e^2 B / 4(2\pi)^2 \hbar$) when $\mu$ lies in the interval between the first Landau levels $n=\pm 1$. The opposite chirality contributes to the transport with the same sign.}
\label{fig_mu}
\end{figure} 
\begin{figure}
\centering
(a) \includegraphics[width=0.45\columnwidth]{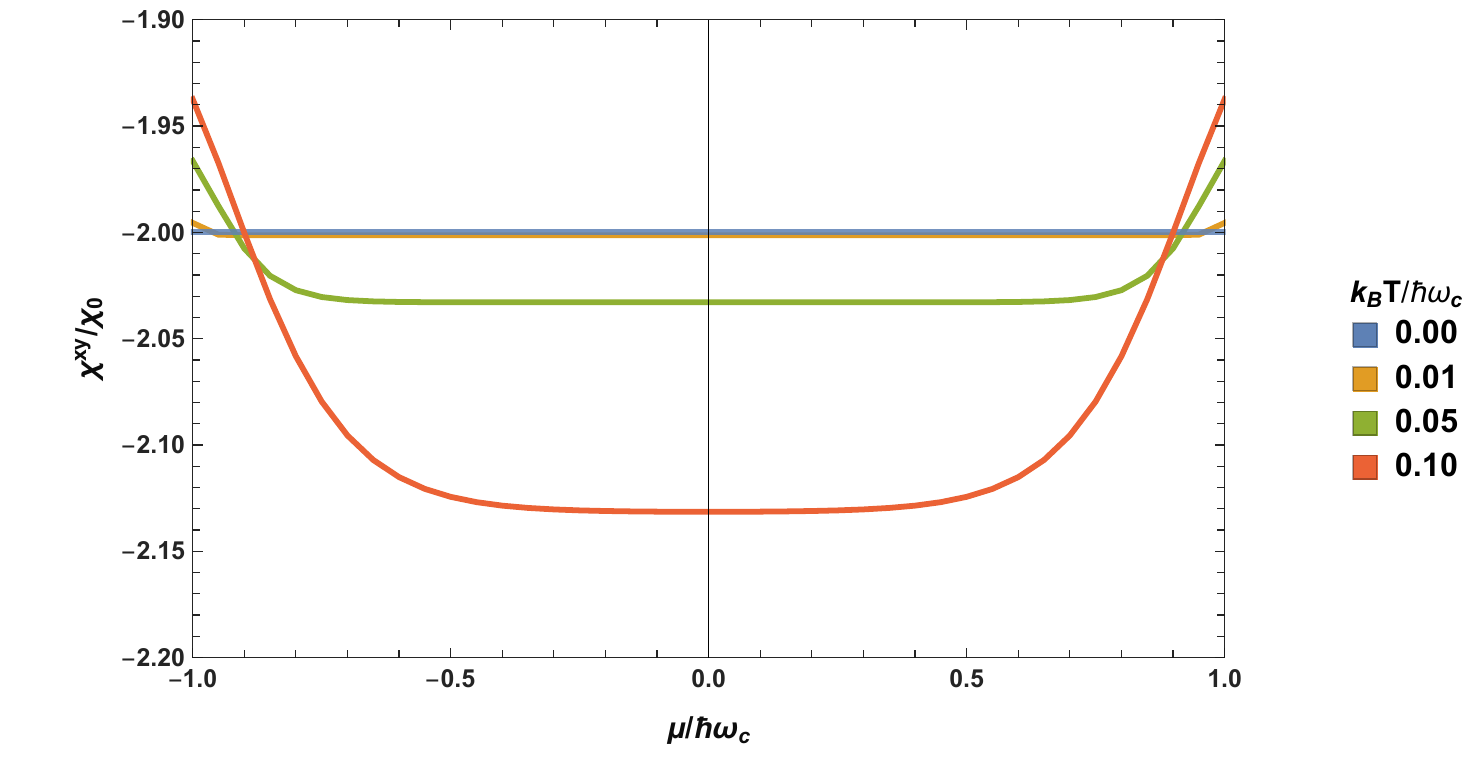} \hspace{0.3cm}
(b) \includegraphics[width=0.45\columnwidth]{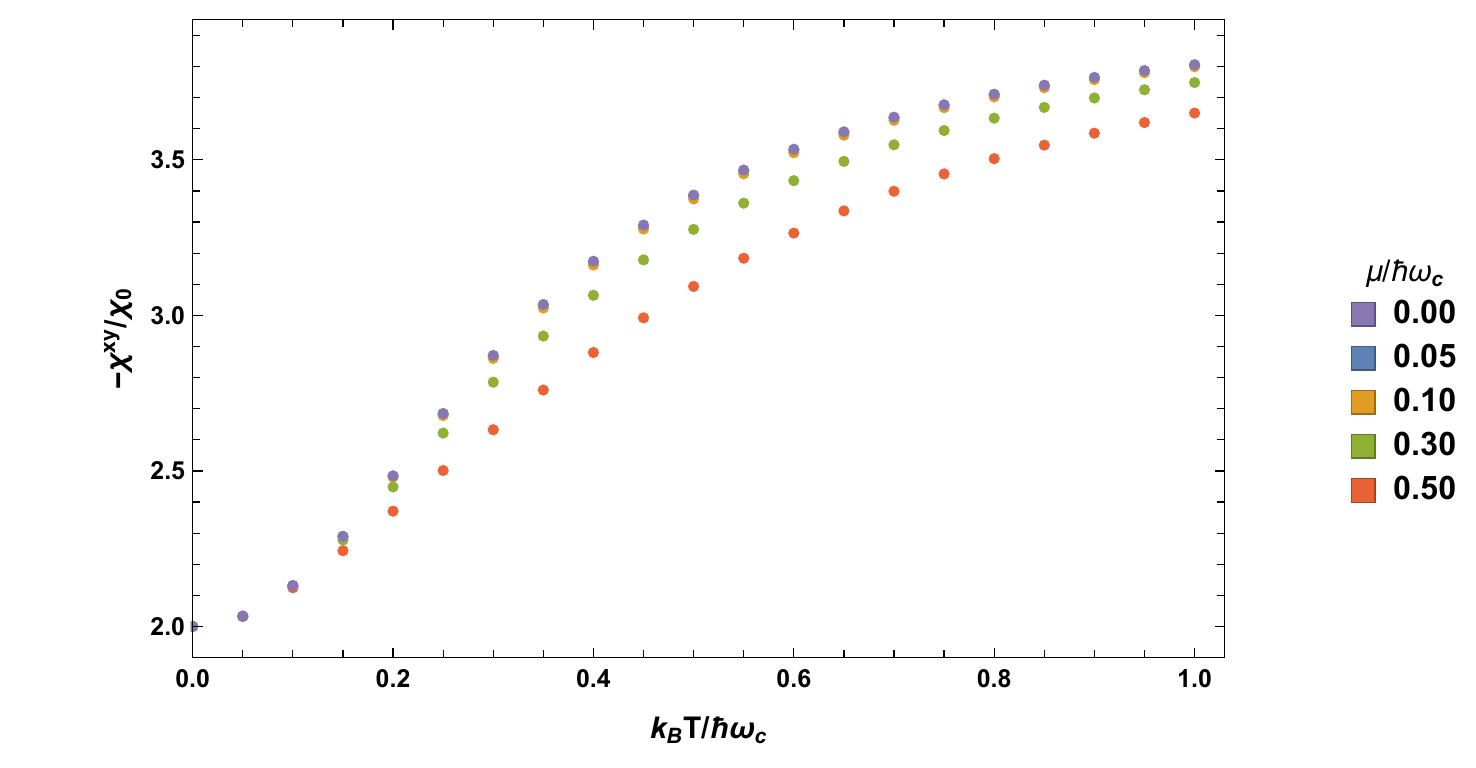}
\caption{(a) Behavior of the thermoelectric coefficient as a function of the chemical potential $\mu$
for  fixed values of $\nabla T/T$. Thermally excited carriers enhance the transport and the boundaries of the plateau are smoothen by the thermal distribution function. (b) Behavior of the thermoelectric coefficient as a function of the temperature for different values of  $\mu$ in the interval $\vert\mu\vert<\hbar \omega_c$.}
\label{fig_muT}
\end{figure} 
To clarify the result and as a guide for the experimentalists we have analyzed the behavior of the thermoelectric coefficient for finite chemical potential $\mu$ and temperature. 
Fig. \ref{fig_mu} shows the Landau level spectrum around a single Weyl chirality of the Dirac semimetal   in normalised units, $\omega_c$ is the cyclotron frequency. The straight line corresponds to the zeroth chiral Landau level. When the chemical potential lies in the interval  $\vert\mu\vert<\hbar\omega_c$ the response function $\chi^{xy}$ maintains the constant value corresponding to the zero doping result. This is a consistency check since the zeroth Landau level has a constant density of states. The contribution from the opposite chirality has the same sign. In Fig. \ref{fig_muT}  (a) we plot the value of the thermoelectric coefficient as a function of the chemical potential in the same range $\vert\mu\vert<\hbar\omega_c$ for different values of the temperature up to  $k_B T= 0.1\; \hbar \omega_c$.  Thermally activated carriers contribute to higher values of the transport coefficient and the size of the plateau (constant value around $\mu=0$) is reduced according to the Fermi Dirac distribution. Fig. \ref{fig_muT}  (b) shows the behavior of the thermoelectric coefficient as a function of the temperature for different values of  $\mu$ in the interval $\vert\mu\vert<\hbar\omega_c$. 


The best materials to explore the physics described in this work would be Dirac or Weyl semimetals with the Fermi level as close as possible to the Dirac point. As discussed along the work,  the calculation has been done for a single chirality and we have checked that the contribution from the other chirality adds up. So the result is valid irrespective of the relative positions of the Weyl cones. The advantage of Dirac semimetals where the two chiralities are superimposed in momentum and energy, is that it minimises the anomalous contribution coming from the Berry phase --
although there will be an anomalous contribution in the presence of a magnetic field \cite{LLetal17} --. As mentioned in \cite{CCV18}, we can also measure the conformal anomaly contribution to the Nernst effect indirectly in magnetic Weyl semimetals (with the Weyl nodes separated in momentum space) by measuring the $S_{xx}$ component of the  thermopower.  
An interesting material to explore the conformal anomaly contribution to thermal transport would be $GdPtBi$, which has the Fermi level near the Dirac point and has recently been used to discuss the mixed gravitational anomaly contribution \cite{Setal18}. 

In the physics of massless Dirac materials, a current to be anomalous normally is the prize to pay for another current to be conserved. In the case of the axial anomaly, we chose to conserve the vector current (and hence the electric charge) and the axial charge to be anomalous. Something similar happens with the conformal anomaly. The Noether current associated to scale invariance, the dilatation current, is $J_D^\mu= T^{\mu\nu}x_\nu$. Assuming energy-momentum conservation, its divergence is then the trace of the energy-momentum tensor. We could have chosen to conserve the dilatation  current at the quantum level at the prize of breaking  energy-momentum conservation. The relevance of the material realization of Weyl physics is the possibility to explore these choices. While the conservation of the electric charge is a natural choice also in the condensed matter context, we sure can play in the matter realizations with energy and momentum conservation and get a deeper insight into the physics of the anomalies.

\begin{acknowledgments}
We thank  A. Cortijo for invaluable help with the calculations and discussion. Useful comments from K. Landsteiner, I. Shovkovy, F. de Juan, and P. Kim are also acknowledged.  This work  has been supported by the PIC2016FR6/PICS07480, Spanish MECD grant FIS2014-57432-P, the Comunidad de Madrid MAD2D-CM Program (S2013/MIT-3007) and within the state assignment of the Ministry of Science and Higher Education of Russia (Grant No. 3.6261.2017/8.9). 
\end{acknowledgments}

\bibliography{Nernst}

\appendix
\vspace{0.3cm}
\begin{center}
{\bf SUPPLEMENTARY MATERIAL}
\end{center}

\section{Projection of the operators on the Landau wave-functions} 
\label{ap_LL}
The Hamiltonian of the Dirac semimetal in a magnetic field is decomposed into the following two Weyl Hamiltonians: 
\beq 
H_s = s \vf  \sigma^i ( p_i + e A_i) , 
\eeq 
where $s=\pm$ denotes the chirality of the node. In the Landau gauge $A_x= -B y$, the spectrum of the system is:  
\beq 
E_{k_z m s}  = \text{sign}(m) \vf \left[ 2 e \hbar B \vert m \vert  + \hbar^2 k_z^2 \right]^{1/2}      \hspace{1cm}  m\in \mathbb{Z}, \; m\neq 0\;,
\eeq 
\beq 
E_{k_z 0 s} = - s \vf \hbar k_z ,
\label{eq_zeroLL}
\eeq 
and the eigenvectors are: 
\beq 
\varphi_{\bb k m s } (\bb r) = \frac{1}{\sqrt{L_x L_z }} \frac{e^{i k_x x} e^{i k_z z}}{\sqrt{\alpha_{k_z m s}^2+1}}e^{-(y-k_x l_B^2)^2/2l_B^2} \begin{pmatrix} 
 \frac{\alpha_{k_z m s}}{\sqrt{2^{M-1} (M -1)! \pi^{1/2} l_B } } H_{M - 1} \left[\frac{y-k_x l_B^2 }{l_B}\right] \\ \frac{1}{\sqrt{2^{M } M ! \pi^{1/2} l_B } } H_{M } \left[\frac{y-k_x l_B^2 }{l_B}\right]
\end{pmatrix} ,
\label{eqn:lwe}
\eeq
with
\beq
\alpha_{k_z m s} =\frac{-\sqrt{2 e B \hbar \vert m \vert }}{E_{k_z m s}/s \vf - \hbar k_z } .
\eeq 
Capital letters refer to the absolute value of Landau levels, $H_m (x)$ are Hermite polynomials, while the factor $\sqrt{\alpha_{k_z m s}^2 +1 } $ comes from the wave-function normalization. The term $\alpha_{k_z m s}$ is chiral dependent only when the Landau level $m$ is different from 0. The field operators are expressed, in the Landau basis, as: 
\begin{alignat}{2}
\hat\psi^\dagger (t,\bb r) &= \sum_{\bb k m} \langle \bb k m s \vert \bb r \rangle \hat a_{\bb k m s}^\dagger (t)= \sum_{\bb km} \varphi^\ast_{\bb k m s} (\bb r) \hat a_{\bb k m s}^\dagger (t)
\label{eqn:lbasis1} \\
\hat\psi (t,\bb r) &= \sum_{\bb l n} \langle \bb r \vert \bb l n s  \rangle \hat a_{\bb l n s} (t)= \sum_{\bb l n} \varphi_{\bb l n s} (\bb r) \hat a_{\bb l n s}(t)
\label{eqn:lbasis2}
\end{alignat}
Using \Cref{eqn:lbasis1,eqn:lbasis2}, the current operator and the energy momentum tensor defined in \Cref{eq:rf} are trivially represented in the Landau basis: 
\beq
J^x(t,\bb r) = s \vf e \sum_{\bb km,\bb l n} \varphi^\ast_{\bb k m s} (\bb r ) \sigma^x \varphi_{\bb l n s} (\bb r)  \hat a^\dagger_{\bb k m s} (t)  \hat a_{\bb l n s}(t),
\eeq 
\begin{alignat}{2}
T^{0y}(t'', \bb r ') &=\sum_{\bold{\boldsymbol{\kappa}}\mu,\bold{\boldsymbol{\lambda}}\nu} \frac{1}{4} \left[ \vf \varphi_{\bold{\boldsymbol{\kappa}} \mu s}^\ast (\bb r ') \mathbb{I}  
\left\{ p_y \varphi_{\bold{\boldsymbol{\lambda}} \nu s}  (\bb r ' )\right \} 
 -\vf \left\{p_y  \varphi_{\bold{\boldsymbol{\kappa}} \mu s}^\ast (\bb r ') \right\}
 \mathbb{I}   \varphi_{\bold{\boldsymbol{\lambda}} \nu s}  (\bb r ' )    \right]
  \hat a^\dagger_{\bold{\boldsymbol{\kappa}} \mu s} (t'')  \hat a_{\bold{\boldsymbol{\lambda}} \nu s} (t'')  \nonumber \\ 
                     &+\sum_{\bold{\boldsymbol{\kappa}}\mu,\bold{\boldsymbol{\lambda}}\nu} \frac{1}{4} \left[ \varphi^\ast_{\bold{\boldsymbol{\kappa}} \mu s}  (\bb r ' ) s \sigma^y \varphi_{\bold{\boldsymbol{\lambda}} \nu s} (\bb r ' ) \right] \left[ \hat a^\dagger_{\bold{\boldsymbol{\kappa}} \mu s} (t'') \{i\hbar \partial_0 \hat a_{\bold{\boldsymbol{\lambda}} \nu s}(t'')\} - \{i\hbar \partial_0 \hat a^\dagger_{\bold{\boldsymbol{\kappa}} \mu s} (t'') \} \hat{a}_{\bold{\boldsymbol{\lambda}} \nu s} (t'' ) \right] \nonumber \\ 
                     &+\sum_{\bold{\boldsymbol{\kappa}}\mu,\bold{\boldsymbol{\lambda}}\nu} \frac{1}{4} \left[ \varphi^\ast_{\bold{\boldsymbol{\kappa}} \mu s}  (\bb r ' ) s \sigma^y \left( 2\mu \right)\varphi_{\bold{\boldsymbol{\lambda}} \nu s} (\bb r ' ) \right] \hat a^\dagger_{\bold{\boldsymbol{\kappa}} \mu s} (t'') \hat a_{\bold{\boldsymbol{\lambda}} \nu s}(t'') ,
\end{alignat}
where, for future purposes, we have introduced a finite chemical potential in the last term of the energy-momentum tensor. All time dependence is given explicitly by the second quantization operators: $\hat a^\dagger_{\bb k m s} (t)= e^{i/\hbar E_{k_z m s} t} \hat{a}^\dagger_{\bb k m s}$.  In momentum space, we get: 
\beq
\hat{J}^x (t, \bb q ) =  \sum_{\bb k m,\bb l n}   J^x_{\bb k m s ,\bb l n s } (\bb q) \hat a^\dagger_{\bb k m s} (t) \hat a_{\bb l n s} (t),
\label{eqn:kk0}
\eeq 
\beq 
\hat{T}^{0y} (t'', -\bb q) =  \sum_{\bold{\boldsymbol{\kappa}} \mu,\bold{\boldsymbol{\lambda}} \nu}  T^{0y}_{\bold{\boldsymbol{\kappa}} \mu s,\bold{\boldsymbol{\lambda}} \nu s} (\bb -\bb q) \hat a^\dagger_{\bold{\boldsymbol{\kappa}} \mu s} (t'') \hat a_{\bold{\boldsymbol{\lambda}} \nu s} (t''),
\label{eqn:kk1}
\eeq 
where the matrix elements are written as a function of the Landau eigenvectors: 
\beq 
J^x_{\bb k m s,\bb l n s } (\bb q) =\frac{1}{(2\pi)^{3/2}} \intt\! \text{d}\bb r \, e^{-i \bb q \bb r} \, s \vf e  \varphi^\ast_{\bb k m s} (\bb r ) \sigma^x \varphi_{\bb l n s} (\bb r) ,
\label{eqn:me1}
\eeq
\begin{alignat}{2}
T^{0y}_{\bold{\boldsymbol{\kappa}} \mu s ,\bold{\boldsymbol{\lambda}} \nu s } (-\bb q ') &= \frac{1}{4} \frac{1}{(2\pi)^{3/2}} \intt \! \text{d}\bb r ' \, e^{+i \bb q \bb r ' }  \left[ \vf \varphi_{\bold{\boldsymbol{\kappa}} \mu s}^\ast (\bb r ') \mathbb{I} \left\{p_y \varphi_{\bold{\boldsymbol{\lambda}} \nu s}  (\bb r ' )\right\}  -\vf \left\{p_y  \varphi_{\bold{\boldsymbol{\kappa}} \mu s}^\ast (\bb r ') \right\} \mathbb{I}   \varphi_{\bold{\boldsymbol{\lambda}} \nu s}  (\bb r ' )    \right] \nonumber \\ 
&+ \frac{1}{4} \frac{1}{(2\pi)^{3/2}} \intt \! \text{d}\bb r ' \, e^{+i \bb q \bb r ' } \left[ \varphi^\ast_{\bold{\boldsymbol{\kappa }\mu s}}  (\bb r ' ) s \sigma^y (E_{\kappa_z \mu s}+E_{\lambda_z \nu s}-2\mu) \varphi_{\bold{\boldsymbol{\lambda}} \nu s} (\bb r ' ) \right] .
\label{eqn:ttoxo}
\end{alignat}
In the chosen gauge, $k_x$ and $k_z$ are still good quantum numbers, and their wave-functions are plane waves (see \Cref{eqn:lwe}). This allows us to establish a relation between the wave-vectors $\bb k$, $\bb l$, $\bold{\boldsymbol{\kappa}}$ and $\bold{\boldsymbol{\lambda}}$. Considering the occupied volume per value, we can replace the summation over $\bb l$ and $\bold{\boldsymbol{\lambda}}$ in \Cref{eqn:kk0,eqn:kk1} by  integrals:
\beq 
\sum_\bb l = \frac{L_x L_y }{4\pi^2} \intt \!\text{d}l_x \text{d}l_z ,
\eeq 
and get two Dirac delta functions correlating the wave-vectors:
\beq 
J^x_{\bb k m s , \bb l n s } (\bb q ) = \frac{1}{(2\pi)^{3/2}} \intt \! \text{d}l_x \text{d}l_z \intt \! \text{d}y \, e^{- i q_y y } \delta(l_x - k_x - q_x ) \delta(l_z - k_z - q_z  ) s \vf e \varphi^\ast_{\bb k m s } (y) \sigma^x \varphi_{\bb l n s }(y) ,
\eeq 
\beq 
T^{0y}_{\bold{\boldsymbol{\kappa}}  \mu s , \bold{\boldsymbol{\lambda}} \nu s } (-\bb q ) = \frac{1}{(2\pi)^{3/2}} \intt \! \text{d}\lambda_x \text{d}\lambda_z  \intt \! \text{d}y' \, e^{ i q_y y' } \delta(\lambda_x - \kappa_x + q_x ) \delta(\lambda_z - \kappa_z + q_z  ) \varphi^\ast_{\bold{\boldsymbol{\kappa}} \mu s } (y') \left( \cdots \right) \varphi_{ \bold{\boldsymbol{\lambda}} \nu s }(y')   ,
\label{eqn:ttoxos}
\eeq 
where $\varphi_{\bb k m s }(y)$ is the remaining part of the wave-function that only depends on $y$:     
\beq 
\varphi_{\bb k m s } (y) =  \frac{1}{\sqrt{\alpha^2_{k_z m s} +1}}e^{-(y-k_x l_B^2)^2/2l_B^2} \begin{pmatrix} 
 \frac{\alpha_{k_z,m,s}}{\sqrt{2^{M-1} (M -1)! \pi^{1/2} l_B } } H_{M - 1} \left[\frac{y-k_x l_B^2 }{l_B}\right] \\ \frac{1}{\sqrt{2^{M } M ! \pi^{1/2} l_B } } H_{M } \left[\frac{y-k_x l_B^2 }{l_B}\right]
\end{pmatrix} .
\eeq
Performing the different integrals one obtains the expressions given in \Cref{eqn:qq2,eqn:qq3} of the main text.

\section{Thermoelectric response}
\label{ap_TR}
The thermoelectric current generated by an external thermal gradient perpendicular to a magnetic field in a Dirac or Weyl semimetal is described, in momentum space, by: 
\beq 
\langle J^x \rangle (\omega , \bb q ) =  \chi^{xy} (\omega, \bb q ) \, (i q_y ) \, g_{00} (\omega , \bb q)  ,
\eeq 
where the response function is written as: 
\beq 
\chi^{xy } (\omega , \bb q) = \lim_{\eta \to 0} \sum_{\bb k ,m n } \frac{(2\pi)^3}{ \mathcal{V}} \frac{-i \vf \hbar J^x_{\bb k m s,\bb k+\bb q ns} (\bb q ) T^{0y}_{\bb k+\bb q n s, \bb k m s} (- \bb q) }{\left (E_{k_zms} - E_{k_z+q_z ns} + i\hbar \eta\right)\left (E_{k_zms} - E_{k_z+q_z ns} + \hbar \omega +i\hbar \eta\right)} \left [  n_{\bb k m s} - n_{\bb k+\bb q ns} \right] .
\eeq
The rest of the calculation consists in computing the numerical value of such expression (once the summation over $\bb k$ has been converted into an integral). The integration over $k_x$ is easily done since the Landau levels are independent of it; only eigenvectors are proportional to this conserved number, the system being degenerated. It can be shown that the product of matrix elements $J^x_{\bb k m s,\bb k+\bb q ns} (\bb q ) T^{0y}_{\bb k+\bb q n s, \bb k m s} (- \bb q) $ does not depend on $k_x$ either, only on the wave-vector $\bb q$. Consequently, the integration over $k_x$ simplifies to add only the degeneracy factor $e B L_y /\hbar$. 

In order to facilitate the calculation, we define a dimensionless variable $\kappa_z = \hbar  k_z/\sqrt{2eB\hbar} $. The WSM eigenvalues are rewritten in the form $E_{k_z m s } = \vf \sqrt{2 e B \hbar}\, \mth E_{\kappa_z m s }$, with $\mth E_{k_z m s}$ a dimensionless energy whose expression depends on the given level. The parameter $\alpha_{k_z m s}$ is also normalized with this change. 

\subsection{Energy-momentum tensor. Product of Hermite polynomials.} 
We will split the  matrix elements of the energy-momentum tensor, defined in \Cref{eqn:ttoxo,eqn:ttoxos}, into three parts: 
\begin{alignat}{2}
T^{0y\,\, [1]}_{\bb k+\bb q n s, \bb k m s} (- \bb q) &= \frac{1}{4} \frac{1}{(2\pi)^{3/2}} \intt \! \text{d} y ' \, e^{+i  q_y y ' }  
\varphi_{\bb k+\bb q n s}^\ast (y') s \sigma^y 
(E_{\kappa_z \mu s}+E_{\lambda_z \nu s}-2\mu)  
\varphi_{\bb k m s }  ( y ' ) \nonumber \\ 
T^{0y\,\, [2]}_{\bb k+\bb q n s, \bb k m s} (- \bb q) &=\frac{1}{4} \frac{1}{(2\pi)^{3/2}} \intt \! \text{d} y ' \, e^{+i  q_y y ' } 
\vf \varphi_{\bb k+\bb q n s}^\ast (y') \mathbb{I} \left\{p_y \varphi_{\bb k m  s}  (y ' )\right\} \nonumber \\ 
T^{0y\,\, [3]}_{\bb k+\bb q n s, \bb k m s} (- \bb q) &=\frac{1}{4} \frac{-1}{(2\pi)^{3/2}} \intt \! \text{d} y ' \, e^{+i  q_y y ' }   
\vf \left\{p_y  \varphi_{\bb k + \bb q n s}^\ast (y') \right\} \mathbb{I}   \varphi_{\bb k m s}  (y' ) 
\label{eqn:emt}
\end{alignat}
It is important to  notice that only the first  expression in \Cref{eqn:emt} depends on the chirality of the node. The operator $p_y$ produces two results when acting on the wave-functions; the first outcome comes from the exponential factor inside $\varphi_{\bb k m s}$. When the same contribution from $T^{0y\,\, [3]}_{\bb k+\bb q n s, \bb k m s}$ is added, a term proportional to $\bb q$ is produced. The resulting expression vanishes at the local limit. The second result appears from the derivative acting on the Hermite polynomials, giving two different terms. Those three contributions will be multiplied by the current operator to provide the thermoelectric response. 

When computing the expressions of the matrix elements \Cref{eqn:me1,eqn:ttoxo}, each current operator is given by the product of Hermite polynomials that satisfy the formula \cite{GR07}: 
\beq 
\intt_{-\infty}^{\infty} \! \text{d} y \, e^{-y^2} H_r (y + a ) H_s (x+b) = 2^s r! \pi^{1/2}  b^{s-r} L_r^{s-r} (-2 a b )  \hspace{1cm} \text{for } s \ge r  ,
\eeq   
where $L_k^{\alpha}(x)$ is the generalized Laguerre polynomial. Depending on the energy levels that we are considering, the position of each polynomial should be modified to satisfy the requirement $s \ge r$, creating three different regimes ($N\le M-1$, $N=M$, $N\ge M+1$) for each matrix product between operators. Taking the local limit, this scenario is greatly simplified, remaining only four contributions to the thermoelectric response function. 

For the sake of clarity, consider as an example the electric current matrix element $J^x_{\bb k m s , \bb k + \bb q n s }$. After performing the spatial integration, the result is: 
\begin{equation}
J^x_{\bb k m s , \bb k + \bb q n s } (\bb q) = \frac{s\vf e }{(2\pi)^{3/2}} \frac{ e^{-( q_x^2+q_y^2) l_B^2/4} \,  \, e^{-i q_y  l_B^2 ( k_x +q_x/2) }}{\left[ \alpha^2_{k_z m s} +1 \vphantom{\alpha^2_{k_z+q_z n s}} \right]^{\frac{1}{2}} \left[ \alpha^2_{k_z+q_z n s} +1 \right]^{\frac{1}{2}}  } \, \Xi_{J^x} (\bb q,m,n,s)\, ,   
\end{equation} 
where the function $\Xi_{J^x} (\bb q , m , n , s)=\Xi_1^{\{i\}}+\Xi_2^{\{j\} }$ encodes all the information related with the different regimes: 
\begin{alignat}{3}
\Xi_1^{\{1\} } (\bb q , m , n , s) & = \alpha_{k_z m s} \sqrt{ \frac{2^N (M-1)!}{2^{M-1} N!} } \left( \frac{- q_x-i q_y  }{2} l_B \right)^{N-M+1} \, L_{M-1}^{N-M+1} \left( \frac{\bb q^2 l_B^2}{2} \right)  &\hspace{1.5cm} (N\ge M-1)  \\ 
\vphantom{.}&\vphantom{.}&\vphantom{.}\nonumber \\ 
\Xi_1^{\{2\} } (\bb q , m , n , s) & = \alpha_{k_z m s} \sqrt{ \frac{2^{M-1} N!}{2^{N} (M-1)!} } \left( \frac{ q_x-i q_y  }{2} l_B \right)^{M-N-1} \, L_{N}^{M-N-1} \left( \frac{\bb q^2 l_B^2}{2} \right)  &\hspace{1.5cm} ( M\ge N+1)  \\ 
\vphantom{.}&\vphantom{.}&\vphantom{.}\nonumber \\ 
\Xi_2^{\{1\} } (\bb q , m , n , s) & = \alpha_{k_z + q_z n s} \sqrt{ \frac{2^{N-1} M! }{2^M (N-1)!} } \left( \frac{- q_x-i q_y }{2} l_B \right)^{N-M-1} \, L_{M}^{N-M-1} \left( \frac{\bb q^2 l_B^2}{2} \right) &\hspace{1.5cm} (N\ge M+1)  \\ 
\vphantom{.}&\vphantom{.}&\vphantom{.}\nonumber \\ 
\Xi_2^{\{2\} } (\bb q , m , n , s) & = \alpha_{k_z + q_z n s} \sqrt{ \frac{2^{M} (N-1)! }{2^{N-1} M!} } \left( \frac{ q_x-i q_y }{2} l_B \right)^{M-N+1} \, L_{N-1}^{M-N+1} \left( \frac{\bb q^2 l_B^2}{2} \right) &\hspace{1.5cm} (M\ge N-1)                        
\end{alignat}
(where $\bb q^2 = q_x^2+q_y^2$). The different pieces that $\Xi$ is made of are chosen depending on the scenario that we are studying. Electric current matrix element will be multiplied with the energy-momentum tensor terms (and their respective $\Xi$ functions). The elements of $\Xi_{J^x}$ and $\Xi_{T^{0y}}$ to be multiplied will be selected according to the regime in consideration. 

\subsection{Thermoelectric response function
}
Computing the product of the electric current with the energy-momentum tensor (the three different parts being defined in the previous section), one gets the expression of the thermoelectric response function \Cref{eqn:rf2}, which is made of two main terms: 
\begin{alignat}{2} 
\lim_{\bb q \to 0} \chi^{xy\,[1]} (\omega, \bb q ) & 
= \frac{1}{ 4(2\pi)^2 } \sum_{\substack{ m,n\\ N=M-1} }  \frac{ \vf  e^2   s^2 B }{\hbar}   \lim_{\eta \to 0}  \intt \! \text{d}\kappa_z \,  \xi (\bold{\boldsymbol{\kappa}}, m , n , s ,\omega,\eta)   (\mth E_{\kappa_z m s} + \mth E_{\kappa_z n s} -2 \mu  ) \, \alpha^2_{\kappa_z m s}
\label{eqn:meh1} \\\nonumber
\vphantom{.}&\vphantom{.}\\
\lim_{\bb q \to 0} \chi^{xy\,[2]} (\omega, \bb q ) 
& = \frac{1 }{4(2\pi)^2 } \sum_{\substack{ m,n\\ N=M-1} } \frac{- \vf e^2 s B }{\hbar}   \lim_{\eta \to 0}  \intt \! \text{d}\kappa_z \,  \xi (\bold{\boldsymbol{\kappa}}, m , n , s ,\omega,\eta)  \left\{ \alpha^2_{\kappa_z m s} \alpha_{\kappa_z n s} \sqrt{M-1}  + \alpha_{\kappa_z m s} \sqrt{M} \right\} ,
\label{eqn:meh2}
\end{alignat}  
where the function $\xi (\bold{\boldsymbol{\kappa}}, m , n , s ,\omega,\eta)$ is defined as: 
\beq 
\xi (\bold{\boldsymbol{\kappa}}, m , n , s ,\omega,\eta) =  \frac{  2 \left [   n_{\bold{\boldsymbol{\kappa}} m s} - n_{\bold{\boldsymbol{\kappa}}  n s} \right] }{\left (\mth E_{\kappa_z m s} - \mth E_{\kappa_z n s} + i\frac{\hbar \eta}{\vf \sqrt{2eB\hbar}}\right)\left (\mth E_{\kappa_z m s} -\mth E_{\kappa_z n s} + \frac{\hbar \omega}{\vf \sqrt{2eB\hbar}} +i\frac{\hbar \eta}{\vf \sqrt{2eB\hbar}}\right)} \frac{1}{\left[\alpha^2_{\kappa_z m s}+1\right] \left[\alpha^2_{\kappa_z n s }+1\right]}.
\eeq 
Some relevant observations on \Cref{eqn:meh1,eqn:meh2}  are the following: as we have seen, the introduction of Dirac delta functions restricts the choices of the transitions between different Landau levels. They arise after computing the product of matrix elements, the result being proportional to $\bb q$ to the ($N\pm M \pm 1$)th power. In order to obtain a non-vanishing result in the local limit ($q\to 0$), the exponents should be zero.  Generalized Laguerre polynomials $L_{k}^{\alpha}(x)$, obtained after the product of Hermite polynomials, are equal to one when the local limit and the Dirac delta functions are evaluated, for any Landau level. Concerning the dependence of the different factors on the chirality we note that although both expressions in \Cref{eqn:meh1,eqn:meh2} are proportional to the chiral factor $s$, the same result is obtained when the other chirality is considered. No cancellations occur and both nodes contribute equally to the response function.

\end{document}